\documentclass[10pt]{article}
\usepackage{graphicx}
\usepackage[T1]{fontenc}
\usepackage[utf8]{inputenc}
\usepackage{amssymb}
\usepackage{amsfonts}
\usepackage{dsfont}
\usepackage{mathtools}
\usepackage{amsthm}
\usepackage{amsmath}
\usepackage{relsize}
\usepackage{textcomp}
\usepackage{eurosym}
\usepackage{stmaryrd}
\usepackage{xcolor}
\usepackage[multiple]{footmisc}
\usepackage{pdflscape}

\usepackage[font=small,labelfont=bf]{caption}

\usepackage{bigints}
\usepackage{geometry}
\begin{document}

\title{Automated Market Makers: Mean-Variance Analysis of LPs Payoffs and Design of Pricing Functions}

\author{Philippe \textsc{Bergault}\footnote{Université Paris Dauphine-PSL, Ceremade, 75116, Paris, France, bergault@ceremade.dauphine.fr.} \and Louis \textsc{Bertucci}\footnote{Institut Louis Bachelier, 75002, Paris, France, louis.bertucci@institutlouisbachelier.org.} \and David \textsc{Bouba}\footnote{Swaap Labs, d@swaap.finance.} \and Olivier \textsc{Guéant}\footnote{Université Paris 1 Panthéon-Sorbonne, Centre d'Economie de la Sorbonne, 106 Boulevard de l'Hôpital, 75642 Paris Cedex~13, France, olivier.gueant@univ-paris1.fr.}}
\date{}

\maketitle
\setlength\parindent{0pt}

\begin{abstract}

With the emergence of decentralized finance, new trading mechanisms called Automated Market Makers have appeared. The most popular Automated Market Makers are Constant Function Market Makers. They have been studied both theoretically and empirically. In particular, the concept of impermanent loss has emerged and explains part of the profit and loss of liquidity providers in Constant Function Market Makers. In this paper, we propose another mechanism in which price discovery does not solely rely on liquidity takers but also on an external exchange rate or price oracle. We also propose to compare the different mechanisms from the point of view of liquidity providers by using a mean / variance analysis of their profit and loss compared to that of agents holding assets outside of Automated Market Makers. In particular, inspired by Markowitz' modern portfolio theory, we manage to obtain an efficient frontier for the performance of liquidity providers in the idealized case of a perfect oracle. Beyond that idealized case, we show that even when the oracle is lagged and in the presence of adverse selection by liquidity takers and systematic arbitrageurs, optimized oracle-based mechanisms perform better than popular Constant Function Market Makers.\\

\medskip
\noindent{\bf Key words:} Automated market makers, cryptocurrencies, DeFi, oracles, stochastic optimal control.\vspace{8mm}

\end{abstract}

\section{Introduction}

Since the early days of the Decentralized Finance (DeFi) era, Automated Market Makers (AMMs) have been some of the largest DeFi protocols on public blockchains. Although it is nothing more than a smart contract on a blockchain, an AMM should be regarded, from a financial point of view, as a liquidity pool of two assets\footnote{There exist AMMs functioning with pools involving more than two assets, but we focus on the two-asset case throughout this paper.} involving two types of agents: liquidity providers (referred to as LPs) and liquidity takers (referred to as LTs or, more simply, traders). LPs supply reserves to the pool, usually in both assets of the pool. In exchange, they become entitled to a share of the pool that is in line with their supply. LTs, in turn, use AMMs to trade, that is to swap a given quantity of one asset against the other. The exchange rate proposed by an AMM is typically based on a pre-defined and public formula that depends on the reserves, the transaction size and the direction of the swap. This function is often called the pricing function, and, although it must be defined in the contract, it can use external data through what is called oracles.\\

AMMs constitute a new paradigm beyond (i) that of dealer markets (like the global FX market, see \cite{schrimpf_fx_2019, schrimpf2019sizing}) where specific agents -- usually banks -- provide liquidity to clients and hold risk on their balance sheet, (ii)~that of classical order-driven markets organized around limit order books (like most stock markets), and (iii) that of dark pools introduced in the last decades (see \cite{lehalle2018market}). Unlike in dealer markets, any agent can provide liquidity through an AMM. Unlike in limit order books, prices are automatically set by the protocol. Unlike in dark pools, the available liquidity is visible and the price is not defined solely by importing that of another venue. Most importantly, LPs do not provide liquidity to LTs in the case of AMMs: LPs provide liquidity to the pool and LTs take liquidity from the pool.\footnote{In particular (like for most DeFi applications) AMMs are fully collateralized and neither LPs nor LTs are exposed to any sort of counterparty risk. To be more precise, LPs are exposed to a technological -- or cyber -- risk, but as long as the smart contract works as expected, they will be able to withdraw their share of the reserves.} The novelty of AMMs raises a lot of theoretical questions from an economic point of view, in relation to the classical market microstructure literature, but also optimization and quantitative questions related to the quantitative finance and financial engineering literature.\\

This paper is a contribution to the quantitative finance literature on AMMs focusing on the relation between the design of AMMs and the performance of LPs. Using standard convex analysis, we recover the now well-known result that, in the absence of transaction fees, posting liquidity into a Constant Function Market Maker\footnote{We refer to \cite{mohan2022automated} for a pedagogical introduction to CFMMs and the classical pricing functions. Examples of popular CFMMs include Uniswap (see \cite{adams_uniswap_2018, adams2020uniswap, adams_uniswap_2021}), Balancer (see \cite{balancer_2019}), Curve (see \cite{curve_2019, curve_2021}), etc. An interesting empirical analysis of Uniswap~v3 is \cite{loesch_impermanent_2021}.} (CFMM) exposes to a concave payoff that is inferior to that of holding coins outside of the pool.\footnote{This led to the now classical concept of impermanent loss that is widely used in the DeFi world. The ``impermanent'' trait of such loss lies in the fact that losses vanish when the exchange rate reverts back to its original value (at the time the liquidity was provided).} This nonpositive and concave payoff comes from the fact that price discovery in CFMMs is left to LTs who act as arbitrageurs and make money out of the pool. Transaction fees should therefore compensate a risk that is fundamentally related to the design of CFMMs and competition between CFMMs cannot reduce transaction fees below a threshold that depends on market conditions. We therefore plead in this paper for the design of oracle-based AMMs that are more complicated than CFMMs in that they do not solely rely on LTs for price discovery.\\

In our paper, we explore indeed AMMs in which the pricing function uses external information about current market prices. Even if a blockchain only has knowledge about on-chain activity, it is possible to feed a smart contract with external data through oracles (see \cite{capponi2023} for a discussion). This situation is quite similar to that of a traditional market maker in dealer markets who provides quotes based on an estimation of the price or exchange rate (typically a mid-price imported from an electronic platform or a composite price) that is available at the same time. Instead of being entirely passive, the AMM can update its bid and offer not only after a trade or a provision or redemption of liquidity, but also after the oracle has fired a price update. One difference with the traditional finance case is however that a dealer in FX or corporate bond markets can have and frequently has short positions whereas liquidity must always be present in the pool in the case of an AMM.\\

In order to compare different AMMs (from the point of view of LPs\footnote{Due to the challenges posed by impermanent loss, much of the existing literature has sought to optimize AMM designs with a focus on LPs. Our work aligns with this perspective, drawing additional inspiration from research works on optimal market making in OTC markets, which predominantly consider the dealers' standpoint. It is worth noting, however, that an economic perspective on AMM design should ideally encompass both LTs and LPs. In the below literature review, there are a few papers tackling equilibrium issues encompassing both LPs and LTs but not with designs like ours.}), we build in this paper a simple mean-variance framework inspired by the so-called modern portfolio theory of the 1950s. However, unlike in the Markowitz case, we always consider the Hodl\footnote{Hodl is slang in the cryptocurrency community for holding a cryptocurrency. In traditional finance, this would be called ``buy and hold''.} strategy as a benchmark. An agent faces indeed an alternative: providing liquidity by posting coins into an AMM or holding them. The Hodl benchmark is also intimately linked to the notion of impermanent loss that is ubiquitous in the field.\\

We also borrow from the modern portfolio theory the concept of efficient portfolios which becomes in our case efficient market making strategies, i.e. efficient pricing functions. Indeed, in addition to comparing the risk / return profile of different existing AMMs, we are interested in computing the maximum extra return that a LP could expect for a given level of tracking error with respect to Hodl.\\

Although the efficient frontier cannot be computed in closed form, it can easily be approximated using a model inspired by the modern literature on market making. Interestingly, we show that popular CFMMs exhibit poor performances relatively to our approximation of the theoretical efficient frontier in a complete information framework. However, even  in the presence of a lagged oracle and adverse selection, an optimized oracle-based AMM is able to get close to that theoretical efficient frontier.\\

In recent years, many research works have been carried out on AMMs, especially CFMMs, with a special focus on Constant Product Market Makers (CPMMs) and their generalizations. Motivated by the emergence of Uniswap and the volume exchanged through it, the authors of \cite{angeris_analysis_2019} studied the main properties of CPMMs. In particular, they showed that, in the presence of arbitrageurs, the exchange rate proposed by a CPMM for small transactions should be in a range around that of competing venues -- the width of the range being determined by transaction fees. Then, they studied the main properties of CPMMs: the no-splitting property, the no-depletion property, etc. They also studied the payoff of LPs in a CPMM as a function of an external asset price when there is no transaction fees (see also \cite{clark_replicating_2020} and an extension to the case of liquidity concentration as in Uniswap~v3 in \cite{clark_replicating_2021}). The returns of LPs are also studied in \cite{boueri2021g3m} and \cite{evans_liquidity_2020} for pricing functions that are geometric means. In particular, impermanent loss is studied for several pricing functions.\\

Beyond CPMMs, CFMMs have been studied by the authors of \cite{angeris_improved_2020} in a general multi-coin setting. They introduced a natural notion of trading set and showed that its convexity is key to obtain desirable properties. In the no-transaction-fee case, they also obtained a formula for the value of the pool that involves Legendre-Fenchel transforms (see also \cite{angeris_replicating_2021} and Section 2 for a discussion). In particular, they showed that  the returns of LPs between two liquidity provision or redemption events suffered from impermanent loss for all CFMMs. The same group of authors with additional coauthors then shed a new light on their previous works in \cite{angeris_constant_2021} by embedding several problems involving CFMMs in a convex setting.\\

A  recent and important work to better understand the payoff of LPs in CFMMs is \cite{milionis2023automated}. Because the value of a CFMM is a concave function of the exchange rate between the two assets, Itô's formula leads to a decomposition\footnote{When the price process is a martingale this is the Doob-Meyer decomposition of the payoff.} of the payoff of LPs into two terms: a stochastic integral corresponding to the payoff of a self-financing strategy and a nonincreasing and nonpositive term. They use this decomposition to claim that, at least theoretically, part of the risk can be hedged away. In particular, if continuous-time hedging with no friction was possible and if an AMM or an LP implemented the hedging strategy, only part of the impermanent loss would remain that they call Loss-Versus-Rebalancing~(LVR).\footnote{In the very recent paper \cite{milionis2023automated2} they added transaction costs and discrete-time trading and obtained asymptotic results with respect to the frequency of blocks.}\\

Beyond the properties of CFMMs and the returns of LPs, several questions have also been addressed in the literature. The question of the optimal fees is discussed in \cite{evans_optimal_2021}. That of the take rate (the proportion of the fees kept by the protocol) is addressed in \cite{fritsch2022economics}. Strategic liquidity provision is the topic of several papers. With a viewpoint rooted into the classical Kyle/Glosten-Milgrom literature on market microstructure, \cite{aoyagi_liquidity_2020} studies liquidity provision in a CPMM with a focus on equilibrium issues. Equilibrium questions are also discussed in \cite{hasbrouck2022need}. The authors of \cite{cartea2022decentralisednew} and \cite{neuder2021strategic} discuss the impact of liquidity concentration (as in Uniswap~v3) on liquidity provision strategies. A microstructural and game-theoretical approach is proposed in \cite{capponi2021adoption} in which the authors discuss, among many interesting topics, the influence of volatility on the adoption of AMMs. The coexistence of limit order books and AMMs and the competition between them are discussed in \cite{aoyagi_coexisting_2021}, \cite{barbon2021quality} and~\cite{lehar_decentralized_2021}. The execution of orders in CFMMs has also been studied in several papers: \cite{angeris_optimal_2022} tackles the static problem of optimal routing across a set of several CFMMs while \cite{cartea2022decentralised} studies a dynamic problem of optimal execution \textit{à la} Almgren-Chriss on a CFMM.\\

In this paper, we go beyond CFMMs and propose an optimized oracle-based AMM. Our ideas and the model we use are inspired by the recent literature on market making. The quantitative literature on market making initially started in the 1980s with the seminal works \cite{ho1981optimal, ho1983dynamics} (see also the papers \cite{amihud1980dealership} and \cite{o1986microeconomics} from the 1980s). It was revived in 2008 in \cite{avellaneda2008high} where the authors use stochastic optimal control tools. Following the latter paper, market making has become an important research strand in quantitative finance over the last decade. The authors of \cite{gueant2013dealing} presented a thorough analysis of the problem introduced in \cite{avellaneda2008high}, and proposed closed-form approximations of the optimal quotes in the case of exponential intensities. Instead of the expected utility framework of \cite{avellaneda2008high}, our model uses the objective function introduced in~\cite{cartea2014buy}. Since then, a lot of new features have been progressively added to the initial models: the impact of parameters ambiguity is studied in \cite{cartea2017algorithmic}, in \cite{bergault2021size} the authors considered a framework with various trade sizes, etc. Also, specific models for different asset classes have been proposed: for stock options in \cite{baldacci2021algorithmic}, for foreign exchange currencies in \cite{barzykin2021algorithmic}, \cite{barzykin2021market} and \cite{barzykin2022dealing}, and for bonds in \cite{gueant2019deep}.\footnote{See the books \cite{cartea2015algorithmic} and \cite{gueant2016financial} for a detailed bibliography on market making.} Our model extends the recent market making models tackling the problem faced by FX dealers, in which exchange rate dynamics are geometric rather than arithmetic, by considering instead of the total profit and loss (PnL) of the liquidity provider, only the extra money (positive or negative) beyond Hodl.\\

The remaining of the paper is organized as follows. In Section \ref{secCFMM}, we introduce notation and recall classical results about the returns of LPs in CFMMs in a concise manner. Section \ref{secOracle} goes beyond the case of CFMMs and derives approximations of efficient pricing functions for an AMM with complete information (in particular a perfect exchange rate oracle). Section \ref{secArb} relaxes the assumptions of Section \ref{secOracle} to be closer to reality by considering misspecification of the parameters, a lagged oracle and adverse selection by arbitrageurs.\\

\section{The payoff of LPs in CFMMs: a primer}
\label{secCFMM}

\subsection{Notation and definition of the protocol}

CFMMs constitute one of the simplest forms of AMMs. In CFMM protocols, the exchange rate for a given pair of coins or tokens is determined by the function of (i) the reserves in the liquidity pool and (ii) the size (and side) of the prospective transaction. In what follows, we recall the functioning of two-currency CFMMs and the now classical analysis of the PnL of LPs in these protocols. It is noteworthy that, because ownership over the CFMM reserves is defined proportionally to the amounts deposited by LPs, one can consider -- without loss of generality -- the special case of a 1-LP-only system.\footnote{As in most papers of the literature, our analysis is valid between any two liquidity provision/redemption events. In other words, we assume that liquidity evolves according to the swaps placed by LTs only.}
\\

In what follows, we shall denote by $(q_t^0)_{t \ge 0}$ and $(q_t^1)_{t \ge 0}$ the two processes for the reserves corresponding respectively to the number of coins of currency~0 and currency~1 in the pool (time $0$ corresponds to the initial posting of reserves). In the case of a CFMM with no transaction fees, the proposed exchange rate is typically such that the quantity $f(q^0_t, q^1_t)$ remains constant\footnote{Of course, in the case of provision/redemption of liquidity, the quantity changes, hence the above footnote.} before and after a trade where the function $f:\mathbb R_+^* \times \mathbb R_+^* \rightarrow \mathbb R$ is typically increasing with respect to both of its variables, reflecting the intent to swap one currency for the other. For what follows, it is more practical to use an alternative formulation based on level sets. We assume that reserves always satisfy the equation $q_t^0 = \psi(q_t^1)$ where the function $\psi : \mathbb R_+^* \to \mathbb R_+^*$ satisfies the following properties:
\begin{itemize}
\item $\psi$ is decreasing -- this reflects that one currency is swapped for the other
\item  $\lim_{q^1 \to 0^+} \psi(q^1) = +\infty$ and $\lim_{q^1 \to +\infty} \psi(q^1) = 0$ -- to prevent depletion of one of the two currencies
\item $\psi$ is strictly convex -- convexity\footnote{The strictness is important only to differentiate the Legendre transform of $\psi$ below.} guarantees the absence of arbitrage opportunities
\item $\psi$ is continuously differentiable -- to simplify the analysis (it is the case in practice).
\end{itemize}

In this setting, if a client wants to sell $\Delta q^1 >0$ coins of currency~1 to the pool at date $t$, he/she will receive  $\Delta q^0$ coins of currency~0 where
$$q_t^0 - \Delta q^0 = \psi(q_t^1 + \Delta q^1), \quad \text{i.e.} \quad \frac{\Delta q^0}{\Delta q^1} = - \frac{\psi(q_t^1 + \Delta q^1) - \psi(q_t^1)}{\Delta q^1}.$$
Symmetrically, if a client wants to buy $\Delta q^1>0$ coins of currency~1 from the pool at date $t$, he/she will pay $\Delta q^0$ coins of currency~0 where
$$q_t^0 + \Delta q^0 = \psi(q_t^1 - \Delta q^1), \quad \text{i.e.} \quad \frac{\Delta q^0}{\Delta q^1} = \frac{\psi(q_t^1 -\Delta q^1) - \psi(q_t^1)}{\Delta q^1}.$$

Of course, the convexity of $\psi$ ensures that $$- \frac{\psi(q_t^1 + \Delta q^1) - \psi(q_t^1)}{\Delta q^1} \le \frac{\psi(q_t^1 -\Delta q^1) - \psi(q_t^1)}{\Delta q^1},$$ thereby precluding the existence of arbitrage opportunities within the pool.

\subsection{No-arbitrage assumption and PnLs}

Assuming there exists at time $t$ an external market exchange rate $S_t$ for currency~1 (in terms of currency~$0$) at which infinitesimal quantities could be traded, we clearly see from the above equations that there would be an arbitrage opportunity at time~$t$ if $S_t$ was not equal to $- \psi'(q_t^1)$. Neglecting the existence of bid-ask spreads even for tiny transactions, we write $S_t = - \psi'(q_t^1)$ and decide to evaluate (in currency~0 terms) any position in currency~1 with exchange rate~$S_t$.\\

In such an idealized setting, the PnL at time $t$ (in currency~0 terms)\footnote{The PnL needs to be accounted in a given currency. We arbitrarily choose currency~0 in what follows.} of the representative LP is therefore
$$\text{PnL}_t = \left(q_t^0 + S_t q_t^1\right) - \left(q_0^0 + S_0 q_0^1\right)$$ while that of the same agent who would not have posted reserves in the pool would be $$\text{PnL}^{\text{Hodl}}_t = \left(q_0^0 + S_t q_0^1\right) - \left(q_0^0 + S_0 q_0^1\right) = (S_t-S_0)q_0^1.$$

To compare those two PnLs, let us define $\psi^*$ the Legendre-Fenchel transform\footnote{We restrict the function to the interior of its domain.} of $\psi$, i.e.
$$\psi^* : p\in \mathbb{R}^*_- \mapsto \sup_{q \in \mathbb{R}^*_+} pq - \psi(q).$$
The maximizer $q$ in the definition of $\psi^*(p)$ is uniquely defined by $\psi'(q) = p$ and it is hence $q_t^1$ when $p=-S_t$. We therefore obtain
$$\psi^*(-S_t) = -q_t^1S_t - \psi(q_t^1) = -q_t^1S_t - q_t^0 \quad \text{and}  \quad \psi^{*'}(-S_t) = q_t^1.$$
In particular, we have,
$$\text{PnL}_t = \psi^*(-S_0) - \psi^*(-S_t) \quad \text{and}  \quad \text{PnL}^{\text{Hodl}}_t = (S_t-S_0)\psi^{*'}(-S_0).$$

Because $\psi^*$ is strictly convex, its graph lies above the tangent line at all points and we have therefore 
$$\text{PnL}_t - \text{PnL}^{\text{Hodl}}_t = \psi^*(-S_0) - \psi^*(-S_t) - (S_t-S_0)\psi^{*'}(-S_0)\le 0,$$ with equality if and only if $S_t = S_0$. This result corresponds to the notion of impermanent loss since the loss vanishes if the exchange rate goes back to its value when reserves were added to the pool.

\subsection{Analysis and discussion}

The above inequality is fundamental as it claims that, under mild assumptions, there is no benefit in providing liquidity through a CFMM with no fees. In other words, a minimal amount of fees is necessary to encourage liquidity provision. In particular, competition between CFMMs cannot decrease fees to zero.\\

The above computations prove that the payoff of a LP in a CFMM with no fees is not only nonpositive but also concave, i.e. similar to that of an option seller. Assuming that $\psi^*$ is twice continuously differentiable and applying Itô's formula as in \cite{milionis2023automated}, we see that
$$\text{PnL}_t - \text{PnL}^{\text{Hodl}}_t = \int_0^t (\psi^{*'}(-S_s)-\psi^{*'}(-S_0)) dS_s - \frac 12 \int_0^t \psi^{*''}(-S_s) d\langle S\rangle_s $$
where $\langle \cdot\rangle $ denoted quadratic variation. The first term can be (theoretically) hedged away through a self-financing strategy with $-(\psi^{*'}(-S_s)-\psi^{*'}(-S_0))$ coins of asset $1$ at time $s$, while the second term, called LVR in \cite{milionis2023automated}, is always nonpositive because $\psi^*$ is convex.\\

Regarding the first term, continuous-time hedging with no friction is a classical theoretical assumption. However, in practice, hedging raises a lot of questions given the high frequency at which trades happen in AMMs: when should we hedge? on which venue? should we cross the spread in limit order books or post limit orders? what about our market impact? etc. Furthermore, hedging requires to trade on another venue and may be complicated to implement inside an AMM: LPs should carry out the hedging process by themselves, all the more since LPs might not want to be hedged at the level of each AMM but rather at the level of their portfolio to benefit from netting effects.  In any case, one can argue that part of the risk could be hedged away, leading therefore to a reduction of the fees charged by CFMMs to LTs in order to compensate LPs. This is an important route for future research in the field.\\

Regarding the second term, it is intrinsically related to the main problem of CFMM protocols. In a CFMM, the price discovery process indeed occurs solely thanks to trades with LTs, and LTs therefore extract value from LPs. To avoid this value extraction, an interesting idea consists in using a price oracle. This is the purpose of this paper.\\

\section{Efficient pricing functions in the complete information case}
\label{secOracle}

\subsection{Modeling framework}

In this section, we consider a filtered probability space $\left( \Omega, \mathcal{F},\mathbb{P}; \mathbb{F}= (\mathcal{F}_{t})_{t\geq 0} \right)$ satisfying the usual conditions.\\

In the previous section, the automated market making protocol did not rely on any external information to propose exchange rates. It indeed proposed exchange rates for various transaction sizes based only on the reserves lying in the pool at a given time. We now consider the theoretical and idealized case where an exogenous market exchange rate is known, for instance the mid-price on a centralized exchange based on a limit order book, like those of Binance, Kraken or Coinbase.
This exchange rate is indicative rather than tradable (equivalently, we assume that the AMM is not able to trade on other venues) even for infinitesimal sizes but we still denote by $(S_t)_{t\in \mathbb R_+}$ the market exchange rate process, stating the value of currency~1 in terms of currency~0. We assume for it the following dynamics:
$$dS_t = \mu S_t dt + \sigma S_t dW_t,$$
where $\mu \in \mathbb R$ is a known deterministic drift, $\sigma > 0$ a known deterministic volatility and $\left(W_t \right)_{t\in \mathbb R_+}$ a standard Brownian motion.\\

To study the PnL of LPs, we need assumptions on the demand of LTs. To build a parsimonious model, we consider as in \cite{barzykin2022dealing} that transaction sizes are labelled in the accounting currency (currency~0) and that transactions are decomposed into two parts: one part corresponding to the market exchange rate and another part corresponding to a markup (that might very rarely be a discount) that is accounted in currency~0, whatever the side of the transaction, for the sake of simplicity. More precisely, if a client wants to buy $z$ coins of currency~0 at time $t$, then $z/S_t$ coins of currency~1 will be asked and $z\delta^{1,0}(t,z)$ out of the total of $z$ coins of currency~0 will not be transferred to him/her. Symmetrically, if a client wants to sell $z$ coins of currency~0 at time $t$, then $z/S_t$ coins of currency~1 will be offered to him/her and $z\delta^{0,1}(t,z)$ extra coins of currency~0 will be asked as a markup.\footnote{$\delta^{0,1}(t,z)$ and $\delta^{1,0}(t,z)$ converted in basis points, could be regarded as ``mid''-to-bid and ask-to-``mid'' in basis points. Everything works indeed almost as if the prices proposed for swapping were respectively $S_t (1-\delta^{1,0}(t,z))$ (bid) and $S_t (1+\delta^{0,1}(t,z))$ (ask). While $S_t$ serves as an indicative price independent of size $z$, the ultimate exchange rate depends on the size (and direction) of the transaction.}\\

We assume that the markups $\left(\delta^{0,1}, \delta^{1,0}\right)$ belong to
\begin{equation}
\begin{split}
\mathcal{A}:= \Bigg\lbrace \delta = \left(\delta^{0,1},\delta^{1,0}\right) : \Omega \times [0,T]& \times \mathbb{R}_{+}^{*} \mapsto \mathbb{R}^{2} \bigg| \delta \text{ is } \mathcal{P} \otimes \mathcal{B}(\mathbb{R}_{+}^{*})\text{-measurable }\\
\qquad \qquad \qquad \qquad &\text{and } \delta^{0,1}(t,z) \wedge \delta^{1,0}(t,z) \geq -C\ \mathbb P \otimes dt \otimes dz \  a.e.  \Bigg\rbrace, \nonumber
\end{split}
\end{equation}
for a given (large) constant $C>0$. Here, $\mathcal{P}$ denotes the $\sigma$-algebra of $\mathbb{F}$-predictable subsets of $\Omega \times[0,T]$ and $\mathcal{B}(\mathbb{R}_{+}^{*})$ denotes the Borelian sets of $\mathbb{R}_{+}^{*}$.\\

In the model, to simplify the analysis, we accumulate these markups in a process $(X_t)_{t \in [0,T]}$, separated from the reserves. Its dynamics is
$$dX_t =  \int_{z \in \mathbb R_+^*} z\delta^{0,1} (t,z) J^{0,1}(dt,dz) + \int_{z \in \mathbb R_+^*} z\delta^{1,0} (t,z) J^{1,0}(dt,dz),$$
with $X_0=0$, where $J^{0,1}(dt,dz)$ and $J^{1,0}(dt,dz)$ are two $\mathbb R_+^*$-marked point processes modelling transactions through which the AMM sells currency~1 and receives currency~0 (for $J^{0,1}(dt,dz)$) and transactions through which the AMM sells currency~0 and receives currency~1 (for $J^{1,0}(dt,dz)$).\\

These marked point processes also allow to write the dynamics of the reserves:
$$dq^0_t =   \int_{z \in \mathbb R_+^*} z \left(J^{0,1}(dt,dz) -  J^{1,0}(dt,dz) \right) \quad \text{and} \quad dq^1_t =  \int_{z \in \mathbb R_+^*} \frac z{S_t} \left(J^{1,0}(dt,dz) -  J^{0,1}(dt,dz) \right).$$

Because we are interested in the PnL of a LP compared to that of an agent who would have held the coins outside of the AMM, we introduce the following two processes:
$$\left(Y^0_t \right)_{t\in \mathbb R_+} = \left((q^0_t-q^0_0) \right)_{t\in \mathbb R_+} \quad \text{and} \quad \left(Y^1_t \right)_{t\in \mathbb R_+} =  \left((q^1_t-q^1_0) S_t \right)_{t\in \mathbb R_+}.$$
Their dynamics are given by:
$$
dY^0_t =   \int_{z \in \mathbb R_+^*}\!\! z \left(J^{0,1}(dt,dz) -  J^{1,0}(dt,dz) \right) \text{ and }
dY^1_t = \mu Y^1_t dt + \sigma Y^1_t dW_t +  \int_{z \in \mathbb R_+^*}\!\!  z \left(J^{1,0}(dt,dz) -  J^{0,1}(dt,dz) \right).$$

We assume that the processes $J^{0,1}(dt,dz)$ and $J^{1,0}(dt,dz)$ have known intensity kernels given respectively by $(\nu^{0,1}_t(dz))_{t\in \mathbb R_+}$ and $(\nu^{1,0}_t(dz))_{t\in \mathbb R_+}$, verifying
$$\nu^{0,1}_t(dz) = \Lambda^{0,1}\left(z, \delta^{0,1} (t,z)\right)\mathds{1}_{\{q^1_{t-}\ge \frac{z}{S_t}\}}m(dz) \quad \text{and} \quad \nu^{1,0}_t(dz) = \Lambda^{1,0}\left(z, \delta^{1,0} (t,z)\right)\mathds{1}_{\{q^0_{t-}\ge z\}}m(dz),$$
where $m$ is a measure (typically Lebesgue or discrete) and $\Lambda^{0,1}$ and $\Lambda^{1,0}$ are called the intensity functions of the processes $J^{0,1}(dt,dz)$ and $J^{1,0}(dt,dz)$ respectively. In the standard literature on market making (see \cite{barzykin2021market} or \cite{bergault2021size}, for instance), these intensity functions (which correspond to the demand curve of LTs) are assumed of the logistic type, i.e.
$$\Lambda^{0,1}(z,\delta)=\lambda^{0,1}(z) \frac{1}{1+e^{\alpha^{0,1}(z) + \beta^{0,1}(z) \delta}} \quad \text{and} \quad \Lambda^{1,0}(z,\delta)=\lambda^{1,0}(z) \frac{1}{1+e^{\alpha^{1,0}(z) + \beta^{1,0}(z) \delta}},\vspace{-0.1cm}$$
where $\lambda^{0,1}(z) m(dz)$ and $\lambda^{1,0}(z) m(dz)$ describe the maximum number of transactions of size in $[z, z+dz]$ per unit of time (the height of the demand curve) and $\alpha^{0,1}(z)$, $\beta^{0,1}(z)$, $\alpha^{1,0}(z)$, and $\beta^{1,0}(z)$ the shape of the demand curve, in particular the sensitivity to the markups. It is noteworthy that indicator functions represent the impossibility for the AMM to propose exchange rates for transactions that cannot occur because reserves are too low in the demanded currency.\\  

The PnL minus the PnL of Hodl at time $T$, hereafter the excess PnL, of a LP is therefore given by\vspace{-0.1cm}
\begin{align}\label{PnL}
X_T + Y^0_T + Y^1_T &= \int_0^T \int_{z \in \mathbb R_+^*} z\delta^{0,1} (t,z) J^{0,1}(dt,dz) + \int_0^T\int_{z \in \mathbb R_+^*} z\delta^{1,0} (t,z) J^{1,0}(dt,dz)\nonumber\\
&\qquad + \int_0^T \int_{z \in \mathbb R_+^*}\!\! z \left(J^{0,1}(dt,dz) -  J^{1,0}(dt,dz) \right) +  \int_0^T \mu Y^1_t dt +\int_0^T \sigma Y^1_t dW_t\nonumber\\
&\qquad +  \int_0^T\int_{z \in \mathbb R_+^*}\!\!  z \left(J^{1,0}(dt,dz) -  J^{0,1}(dt,dz) \right)\nonumber\\
&= \int_0^T \int_{z \in \mathbb R_+^*} z\delta^{0,1} (t,z) J^{0,1}(dt,dz) + \int_0^T\int_{z \in \mathbb R_+^*} z\delta^{1,0} (t,z) J^{1,0}(dt,dz)\nonumber\\
&\qquad +  \int_0^T \mu Y^1_t dt + \int_0^T \sigma Y^1_t dW_t.
\end{align}

We see that the excess PnL can be decomposed into two parts: one corresponding to the accumulated markups and the other one representing the deviation from the Hodl strategy reserves. In particular, it contains jump terms and a Brownian term.\footnote{As in the recent article \cite{milionis2023automated}, one can argue that the term $\int_0^T \mu Y^1_t dt + \int_0^T \sigma Y^1_t dW_t = \int_0^T (q^1_t - q^1_0) dS_t$ could be hedged, at least theoretically.}\\

\subsection{Towards an efficient frontier}

We now derive the optimal strategy in this framework with complete information for an objective function that is not exactly a mean-variance one but a more practical one for us in order to use the tools of stochastic optimal control theory. In fact, we only consider the variance of $\int_0^T \sigma Y^1_t dW_t$ which is a very reasonable proxy for the variance of the excess PnL, as most of the risk comes from the Brownian term and not from the drift or the jump terms.\\ 

More precisely, for each $\gamma >0$,\footnote{The parameter $\gamma$ can be interpreted in two ways. One can see it as a risk aversion parameter or as a Lagrange multiplier, like in the Markowitz framework.} we introduce the following stochastic optimal control problem:

$$\sup_{(\delta^{0,1},\delta^{1,0}) \in \mathcal A} \mathbb{E}\Bigg[\int\limits_{0}^{T}\Bigg\lbrace \int_{z \in \mathbb R_+^*} \Big(z\delta^{0,1}(t,z)  \Lambda^{0,1}(z,\delta^{0,1}(t,z))\mathds{1}_{\{q^1_{t-}\ge \frac z{S_t}\}} $$$$\qquad \qquad \qquad \qquad+z\delta^{1,0}(t,z)  \Lambda^{1,0}(z,\delta^{1,0}(t,z))\mathds{1}_{\{q^0_{t-}\ge z\}}  \Big)m(dz)+  \mu Y^1_t  - \frac{\gamma}{2} \sigma^2 (Y^1_t)^2  \Bigg\rbrace dt  \Bigg].$$ 
\vspace{0.3cm}

This stochastic optimal control problem has $4$ state variables and is therefore hardly tractable, even numerically. Nevertheless, it is noteworthy that for moderate values of $\mu$, the quadratic penalty provides an incentive to keep the composition of the pool close to the initial one.\footnote{In pratice, the choice of $\gamma$ should be contingent on the pool size and the level of liquidity. For example, a higher $\gamma$ value might be suitable when the pool size is smaller and/or liquidity demand is high. Conversely, a reduced $\gamma$ could be adopted when the pool is more substantial, and liquidity demand is lesser. Notably, the value of $\gamma$  could be adjusted at each time of liquidity provision/redemption. In this paper, as in most of the literature, we analyse the PnL of LPs between two times of liquidity provision/redemption and $\gamma$ is fixed.} Therefore, the no-depletion constraints (which translated into indicators in the above formula) can be regarded as superfluous, and we subsequently approximate the problem by removing the latter terms.\footnote{In fact, in our numerical examples, we observe that, with the markups obtained using this approach, the reserves remain positive at all times, i.e. the constraints are not binding.} In other words, we consider the following modified objective function:

\begin{align*}
\mathbb{E}\Bigg[\int\limits_{0}^{T}\Bigg\lbrace \int_{z \in \mathbb R_+^*} \Big(z\delta^{0,1}(t,z)  \Lambda^{0,1}(z,\delta^{0,1}(t,z))+z\delta^{1,0}(t,z)  \Lambda^{1,0}(z,\delta^{1,0}(t,z))  \Big)m(dz) +  \mu Y^1_t  - \frac{\gamma}{2} \sigma^2 (Y^1_t)^2  \Bigg\rbrace dt  \Bigg].\nonumber
\end{align*}
It is noteworthy that this new problem only depends on one state variable, $Y^1$, whose dynamic is Markovian. It can be addressed with classical tools of stochastic optimal control. For that purpose, we introduce the value function $\theta:[0,T]\times \mathbb R \rightarrow \mathbb{R}$ associated with this stochastic optimal control problem. It is well known that $\theta$ solves the following Hamilton-Jacobi-Bellman equation:\footnote{If continuous-time hedging with no friction was possible as in~\cite{milionis2023automated}, the HJB equation would be Eq. \eqref{eqn:HJB} with $\mu=0$ and $\sigma=0$. This would lead to $\theta(t,y) = C(T-t)$ for some constant $C$ independent of $y$ and an optimal strategy independent of $Y^1_t$. A softer way to consider the possibility to hedge could be, as in \cite{barzykin2021algorithmic}, to add a term $\sup_v v\partial_y \theta - L(v)$ with $L : v \mapsto \psi|v| + \eta |v|^{1+\phi}$ an execution cost function as in the optimal execution literature. In particular, the cost of liquidity and the need to cross the spread would be modeled.}
\begin{equation}
\begin{cases}
 \!&0 = \partial_t \theta(t,y) + \mu y \left( 1 + \partial_y \theta(t,y) \right) - \frac{\gamma}{2} \sigma^2 y^2 + \frac 12 \sigma^2 y^2 \partial^2_{yy} \theta(t,y)\\
\!& \qquad + \text{\scalebox{0.6}[1]{$\bigint$}}_{\!\!\mathbb{R}_{+}^{*}} \left(zH^{0,1} \left(z,\frac{\theta(t,y) -  \theta(t,y- z) }{z}\right) + zH^{1,0} \left(z,\frac{\theta(t,y) -  \theta(t,y+z) }{z}\right) \right)m(dz),\\
\!&\theta(T,y) = 0,
\end{cases}
\label{eqn:HJB}
\end{equation}
where
$$
H^{0,1}:(z,p)\in\mathbb R_+^* \times \mathbb{R} \mapsto \underset{\delta\ge -C }{\sup}\ \Lambda^{0,1}(z,\delta)(\delta-p) \text{ and } H^{1,0}:(z,p)\in\mathbb R_+^* \times \mathbb{R} \mapsto \underset{\delta\ge -C }{\sup}\ \Lambda^{1,0}(z,\delta)(\delta-p).$$

Using the same ideas as in \cite{gueant2017optimal}, we see that for $i\neq j \in \{0,1\}$, the supremum in the definition of $H^{i,j}(z,p)$ is reached at a unique $\bar \delta^{i,j}(z,p)$ given by
$$\bar \delta^{i,j}(z,p) = (\Lambda^{i,j})^{-1} \left(z, -\partial_p{H^{i,j}} (z,p)  \right)$$
where for all $z$, $(\Lambda^{i,j})^{-1}(z, .)$ denotes the inverse of the function $\Lambda^{i,j}(z,.)$. Moreover, the markups that maximize our modified objective function are obtained in the following form
\begin{align}\label{optquotes01}
 \delta^{0,1*}(t,z) = \bar \delta^{0,1} \left(z, \frac{\theta(t,Y^1_{t-}) -  \theta(t,Y^1_{t-}-z) }{z} \right)
 \end{align}
 and
\begin{align}\label{optquotes10}
\delta^{1,0*}(t,z) = \bar \delta^{1,0} \left(z, \frac{\theta(t,Y^1_{t-}) -  \theta(t,Y^1_{t-}+z) }{z} \right).   
 \end{align}

In practice, we proceed in two steps in order to solve the above stochastic optimal control problem. First, we approximate numerically the solution to the HJB equation \eqref{eqn:HJB}. For that purpose, we employ operator splitting in order to deal separately with the differential terms and the non-local terms. For the differential terms, we used an implicit scheme with Neumann conditions at the boundaries ($\pm \bar{y}$ for $\bar{y}$ chosen large). As for the non-local terms, we used a discrete measure $m$ and applied a Newton-Raphson algorithm to resolve the implicit scheme at each time step (we assume that the AMM does not accept trades that go beyond the boundaries $\pm \bar{y}$). This approach proved to be computationally efficient. Once the value function $\theta$ is approximated, the second step consists in plugging the approximation of $\theta$ into Equations \eqref{optquotes01} and \eqref{optquotes10} to obtain the associated markups.\\

\subsection{Numerical example}

In what follows, we are going to illustrate the excess PnL associated with various strategies. In order to illustrate our findings, we use throughout the paper a market simulator based on a discrete-time version of the above framework, with the following realistic parameters:
\begin{itemize}
    \item currency~0: USD, currency~1: ETH
    \item Initial exchange rate: 1600 USD per ETH
    \item Drift: $\mu = 0 \text{ day}^{-1}$
    \item Volatility: $\sigma = 1 \text{ year}^{-\frac 12}$ 
    \item Single transaction size: 4000 $\text{USD}$ (i.e. $m$ is a Dirac mass)
    \item Intensity functions: $\lambda^{0,1} = \lambda^{1,0} = 100 \text{ day}^{-1}$, $\alpha^{0,1} = \alpha^{1,0} = -1.8$, $\beta^{0,1} = \beta^{1,0} = 1300 \text{ bps}^{-1}$ (this corresponds to an average of 86 trades per day when the proposed exchange rate always equals the market exchange rate, 96 trades per day when the proposed exchange rate is the market exchange rate improved by 10 bps, and 62 trades per day when the proposed exchange rate is the market exchange rate worsen by 10 bps)
    \item Initial inventory: $2000000$ USD and $1250$ ETH
    \item Time horizon: $T = 0.5 \text{ day}.$\\
\end{itemize}

\begin{figure}[h]
\centering
\includegraphics[width=\textwidth]{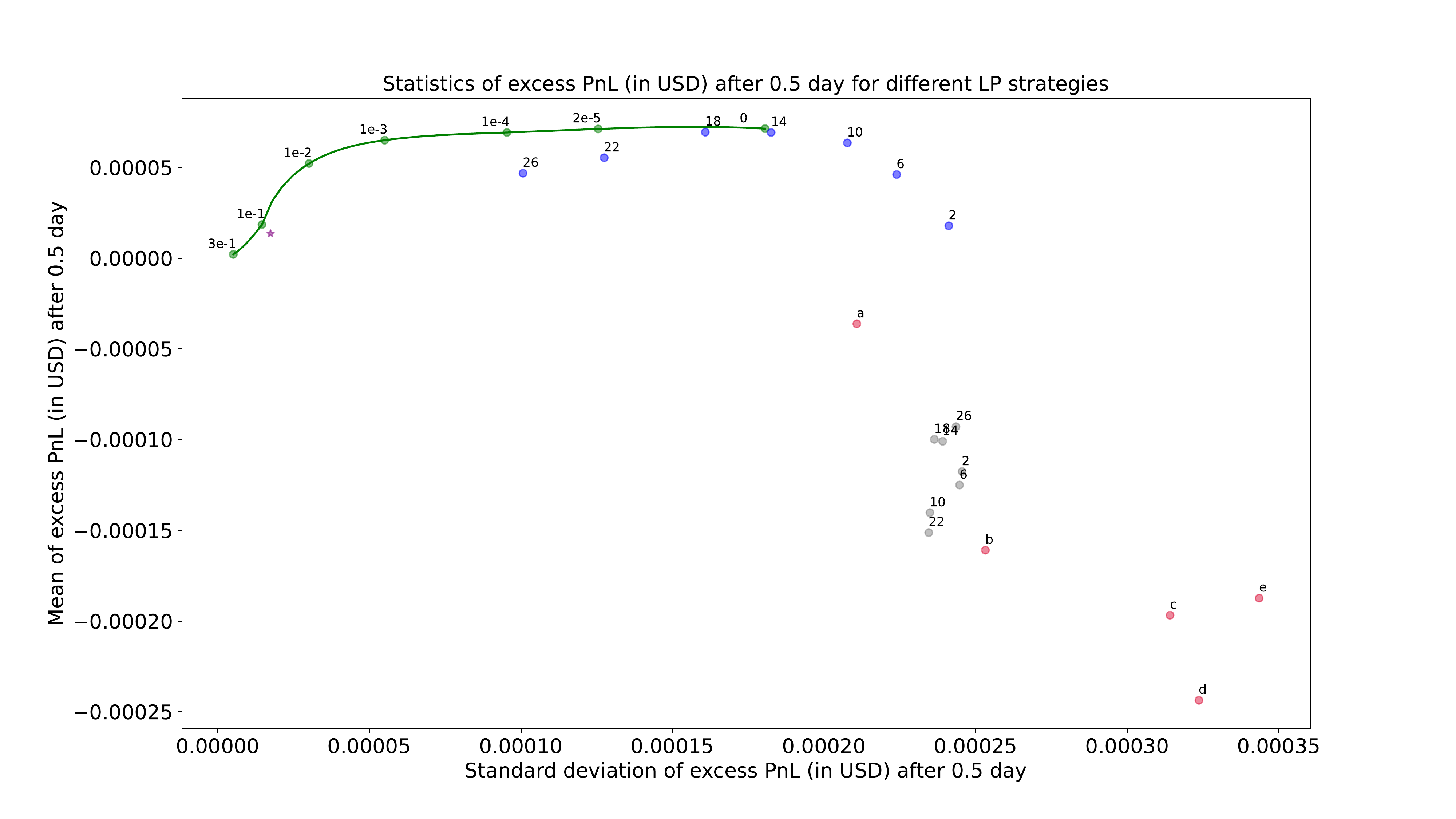}\\
\caption{Performance of strategies in terms of mean / standard deviation of excess PnL. In blue: naive strategies with constant $\delta^{0,1}, \delta^{1,0}$ (the number next to the point corresponds to the value of $\delta^{0,1}$ and $\delta^{1,0}$ in bps). In grey: CPMM with fees (the numbers next to the points correspond to the transaction fees in bps). In pink: CFMM without market exchange rate oracle as described in \cite{curve_2019, curve_2021} for different sets of realistic parameters. In purple~$(\star)$: AMM with market exchange rate oracle as described in \cite{bouba_2021}. In green: approximation of the efficient frontier, obtained using the optimal markups for different levels of risk aversion (the numbers next to the points correspond to the value of $\gamma$).}
\label{pic_1}
\end{figure}

We plot in Figure \ref{pic_1} the performance of AMMs following different strategies:

\begin{itemize}
\item Naive oracle-based strategies consisting in choosing $\delta^{0,1}$ and $\delta^{1,0}$ constant
\item CPMM strategies with various transaction fees
\item CFMM strategies as decribed in \cite{curve_2019, curve_2021}
\item An oracle-based strategy documented in \cite{bouba_2021}
\item Oracle-based strategies associated with the markups we derived from the above stochastic optimal control problem.\\

\end{itemize}

We used Monte-Carlo simulations with 1000 trajectories. The (approximation of the) efficient frontier is parameterized by $\gamma$. Unsurprisingly, when $\gamma$ is large (i.e. large risk aversion), the optimal strategy consists in providing liquidity at a very high cost, and the resulting risk / return profile gets close to the origin $(0,0)$. As $\gamma$ decreases, the efficient frontier describes an increasing curve that stops (unlike in the Markowitz case) at $\gamma = 0$, at a point corresponding to (optimized) constant markups and to the maximum expected excess PnL.\\

Although this first graph corresponds to an idealized world where the market exchange rate is perfectly known at all times (and where there is therefore no arbitrage), it sheds light on the intrinsic limitations of (oracle-free) CFMMs compared to oracle-based AMMs. In our setting, CFMMs indeed underperform optimal strategies, as expected, but also naive strategies, both in terms of expected excess PnL and in terms of standard deviation. It is noteworthy that implementing a hedging strategy for CFMMs would reduce the standard deviation but leave negative expected excess PnLs in our setting.\\

These results, however need to be qualified. Naive strategies with constant transaction fees lack of robustness: in the presence of an arbitrage opportunity due to information asymmetry or mispricing, the pool would indeed be depleted from one of the two assets! In the next section, we show that our optimized oracle-based strategies are, however, robust to misspecifications, lags in price oracles and the introduction of adverse selection.\\

\section{Beyond the idealized case: misspecification, incomplete information and introduction of adverse selection}
\label{secArb}

In practice, an AMM can be designed with some values of the drift, volatility and liquidity parameters, but different values may realize. Furthermore, the drift is highly unpredictable, volatility is not constant but clustered, while liquidity varies depending on market conditions.\footnote{For models taking account of the stochastic nature of volatility and liquidity, see \cite{bergault2023enhancing}.} Consequently, it is of the utmost importance to study the impact of parameters misspecification on the risk~/~return profile of allegedly efficient strategies, i.e. what happens if the realized drift, volatility and liquidity parameters do not match those used in the smart contract.\\

\begin{figure}[h!]
\centering
\includegraphics[width=0.98\textwidth]{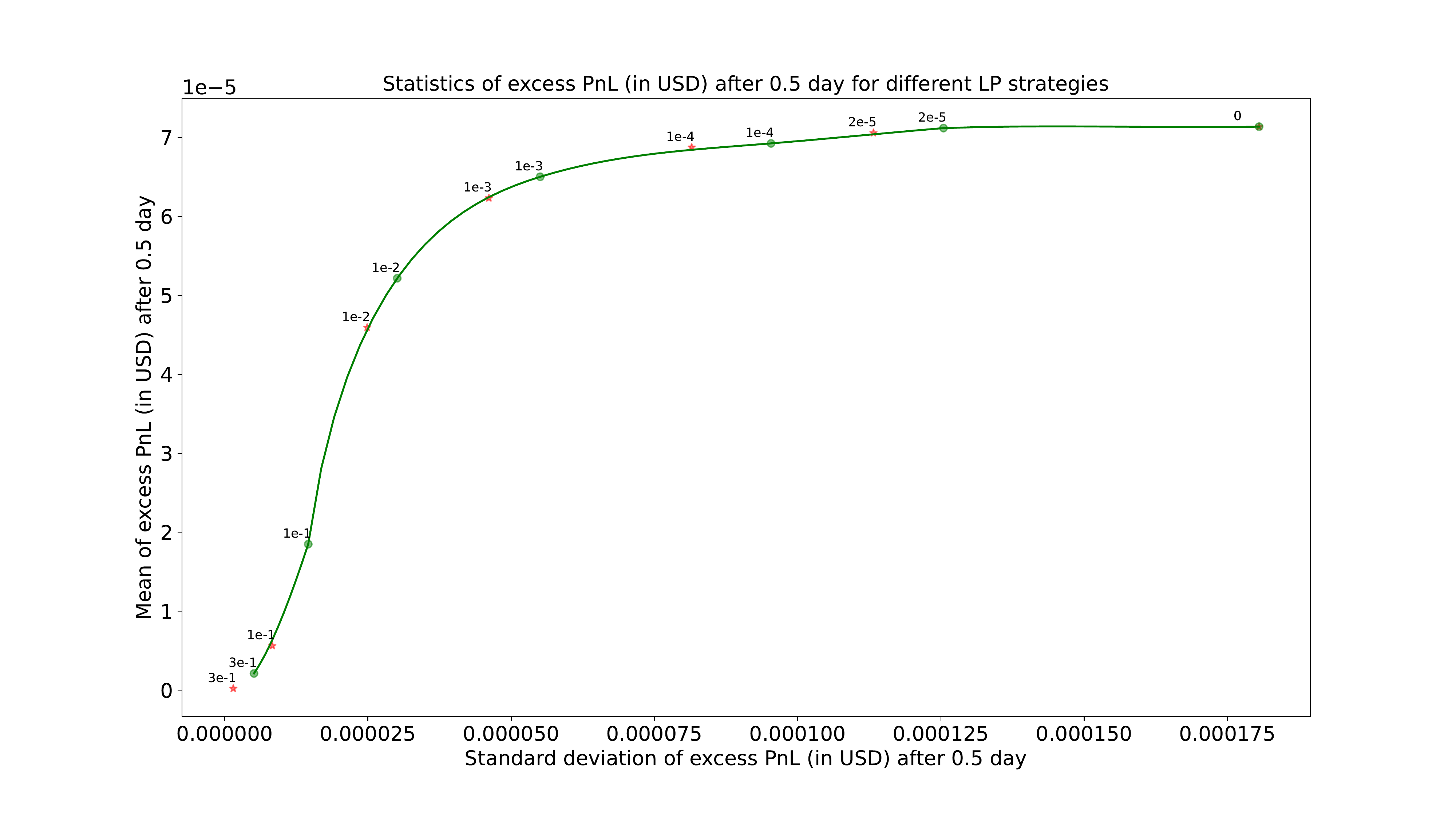}\\
\caption{Performance of strategies in terms of mean / standard deviation of excess PnL when $\lambda^{0,1}$ and $\lambda^{1,0}$ are misspecified. In green: efficient frontier, obtained with the efficient strategy for different levels of risk aversion with complete information. In pink: performance of the misspecified strategy obtained with $\lambda^{0,1} = \lambda^{1,0} = 50\text{ day}^{-1}$ for different levels of risk aversion. The numbers next to the points correspond to the value of $\gamma$.}
\label{pic_mis_lambda200}
\end{figure}

\begin{figure}[h!]
\centering
\includegraphics[width=0.98\textwidth]{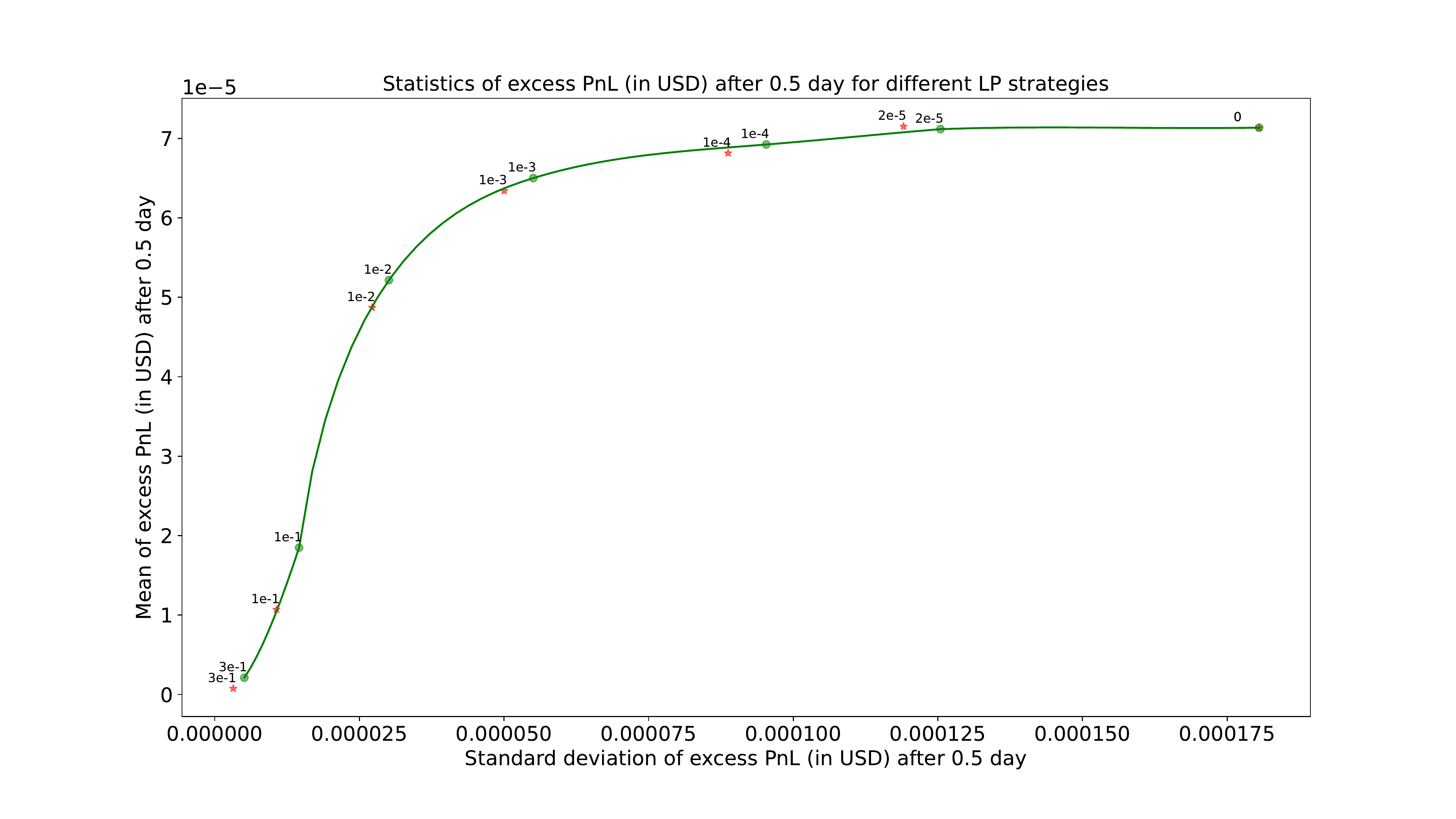}\\
\caption{Performance of strategies in terms of mean / standard deviation of excess PnL when $\sigma$ is misspecified. In green: efficient frontier, obtained with the efficient strategy for different levels of risk aversion with complete information. In pink: performance of the misspecified strategy obtained with $\sigma = 1.2\text{ year}^{-\frac 12}$ for different levels of risk aversion. The numbers next to the points correspond to the value of $\gamma$.}
\label{pic_mis_sigma70}
\end{figure}

In Figure \ref{pic_mis_lambda200}, we consider the case where the actual liquidity parameters $\lambda^{0,1}$ and $\lambda^{1,0}$ one inputs in the simulator are the same as in the above section, but the liquidity parameters of the strategy are set to $\lambda^{0,1} = \lambda^{1,0} = 50\text{ day}^{-1}$. We compare the results to the efficient frontier and observe that the misspecified strategies remain almost exactly on the efficient frontier, with a shift toward higher risk aversions. Similarly, we consider in Figure \ref{pic_mis_sigma70} the case where the actual volatility one inputs in the simulator is equal to~100\% ($\sigma = 1 \text{ year}^{-\frac 12}$) as above, but the volatility used to compute the strategy is 120\% ($\sigma = 1.2 \text{ year}^{-\frac 12}$). We compare the results to the efficient frontier and observe the same phenomenon as for liquidity parameters. These results can be explained theoretically. Indeed, in the absence of drift and ignoring the Laplacian term $\frac 12 \sigma^2 y^2 \partial^2_{yy} \theta(t,y)$, it can be shown that when $\lambda^{0,1} = \lambda^{1,0} =: \lambda$, misspecifying $\lambda$ or $\sigma$ has the same effect as choosing another $\gamma$, because the solution to Equation \eqref{eqn:HJB} depends only on the ratio $\frac{\lambda}{\gamma \sigma^2}$. This is in line with the observation made in \cite{barzykin2021market}.\\

\begin{figure}[h!]
\centering
\includegraphics[width=\textwidth]{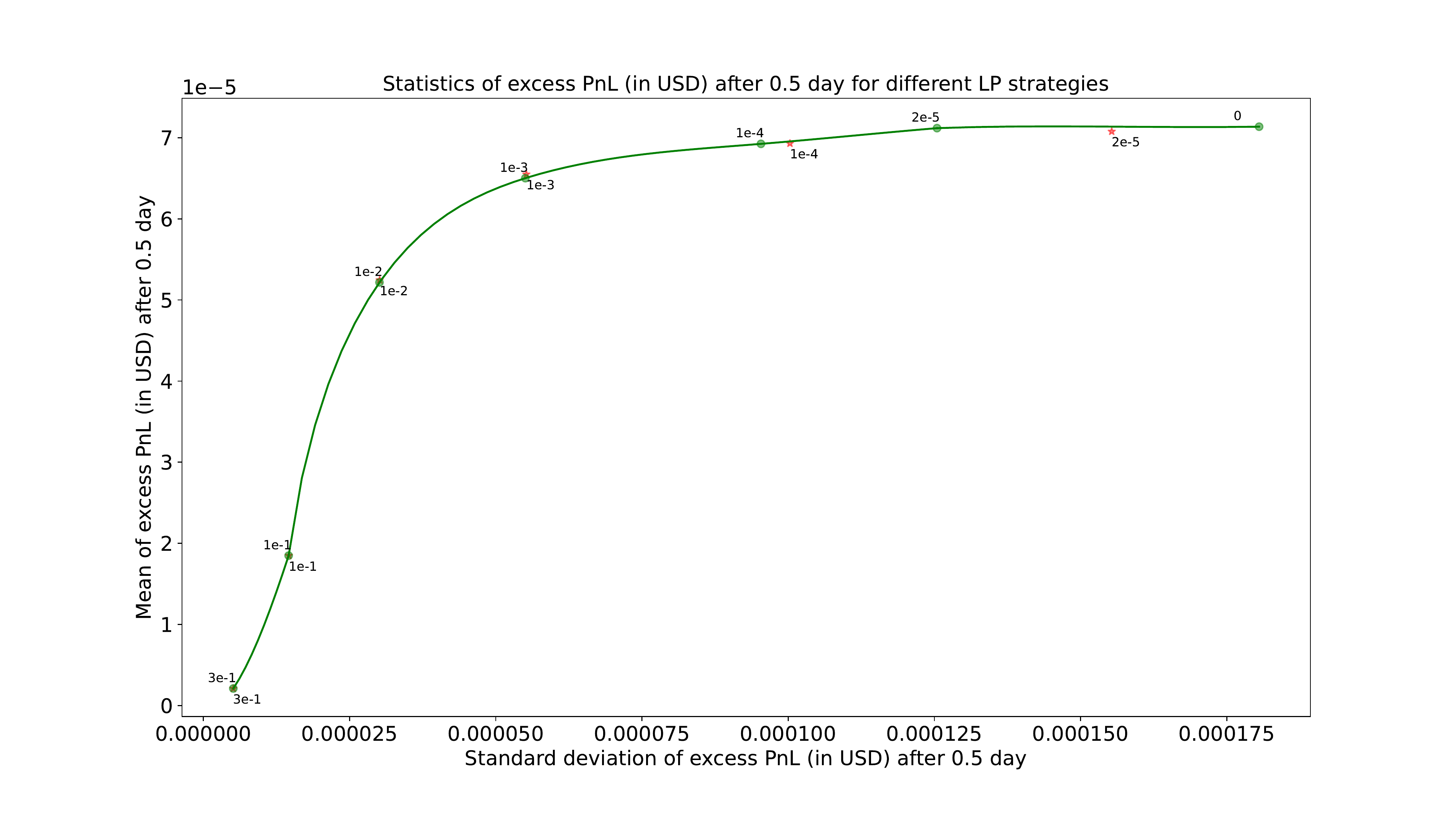}\\
\caption{Performance of strategies in terms of mean / standard deviation of excess PnL when $\mu$ is misspecified. In green: efficient frontier, obtained with the efficient strategy for different levels of risk aversion with complete information. In pink: performance of the misspecified strategy obtained with $\mu = 0.4 \text{ year}^{-1}$ for different levels of risk aversion. The numbers next to the points correspond to the value of $\gamma$.}
\label{pic_mis_mu40}
\end{figure}

Finally, we consider in Figure \ref{pic_mis_mu40} the case where the parameters of the strategy correspond to those used in the previous section with no drift, but the  drift one inputs to compute the strategy is equal to 40\% ($\mu=0.4\text{ year}^{-1}$). We compare the results to the efficient frontier and observe that the misspecified strategies remain close to the efficient frontier. It is noteworthy that we did not compute the point corresponding to no risk aversion ($\gamma = 0$) because the optimal strategy of any risk-neutral agent with a view on the drift is to get an infinite directional position: here, this means that the AMM should sell one of the two assets to buy the other.\\

\begin{figure}[h]
\centering
\includegraphics[width=0.98\textwidth]{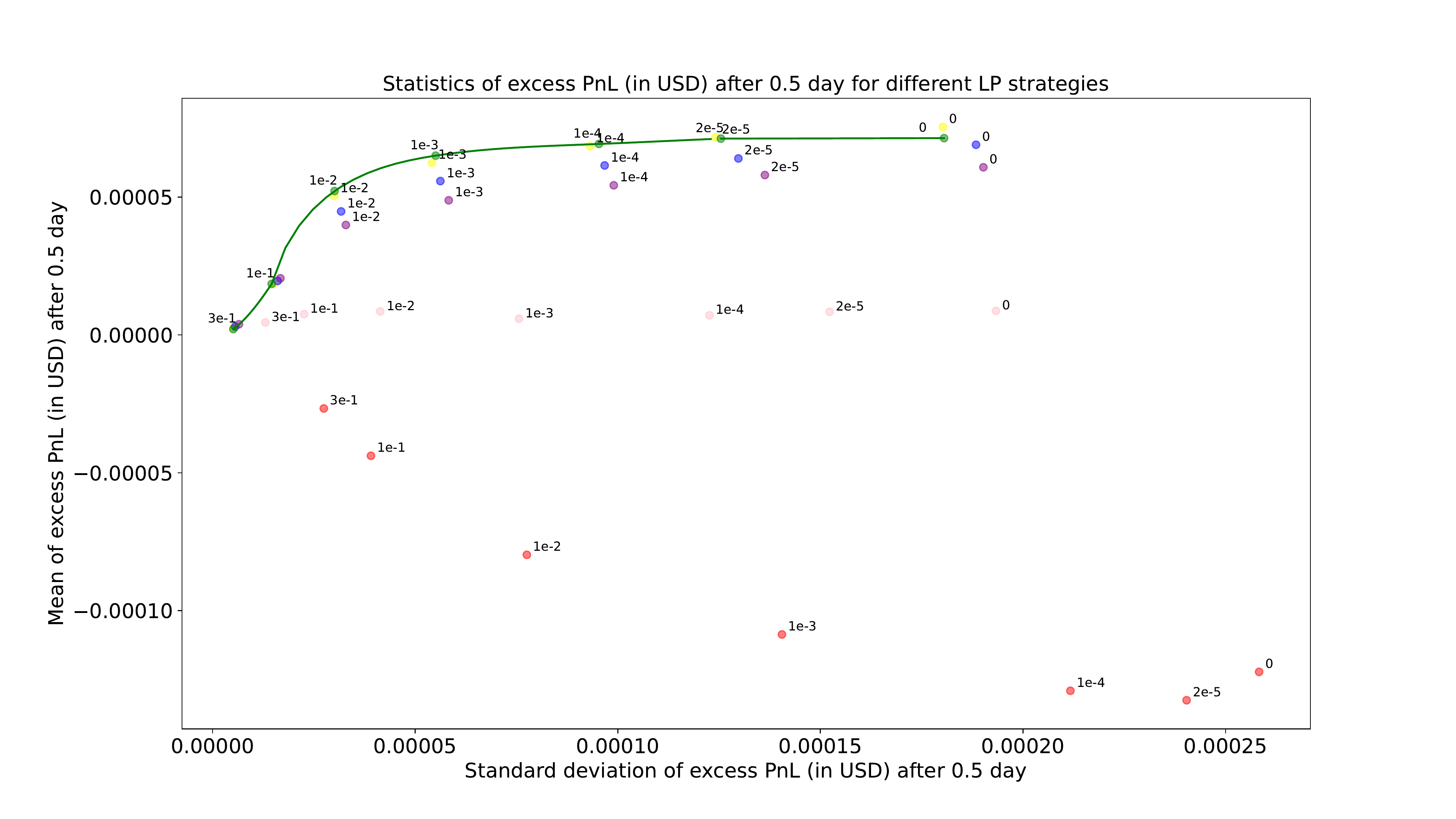}\\
\caption{Performance of the efficient strategies in terms of mean / standard deviation of excess PnL, obtained by playing the efficient strategies for different levels of risk aversion with different oracle delays: complete information (in green), 10-second delay (in yellow), 30-second delay (in blue), 1-minute delay (in purple), 5-minute delay (in pink), 30-minute delay (in red). The number next to the point corresponds to the value of $\gamma$. }
\label{pic_2}
\end{figure}

Misspecification can be a problem, but the main problem faced by AMMs is adverse selection. It is important to recall that the main problem of CFMMs is that value is extracted by LTs. Of course, in the above model with complete information, the market exchange rate is known at all times and adverse selection does not exist. To introduce adverse selection, we assume that the AMM cannot observe the market exchange rate perfectly but rather with a lag, while the demand curves of LTs are centered on the right (current) price. We show the results in Figure \ref{pic_2} and clearly see that performances move away from the efficient frontier as the delay increases. However, the performance remains good for reasonable values of the lag.\\

\begin{figure}[h!]
\centering
\includegraphics[width=0.98\textwidth]{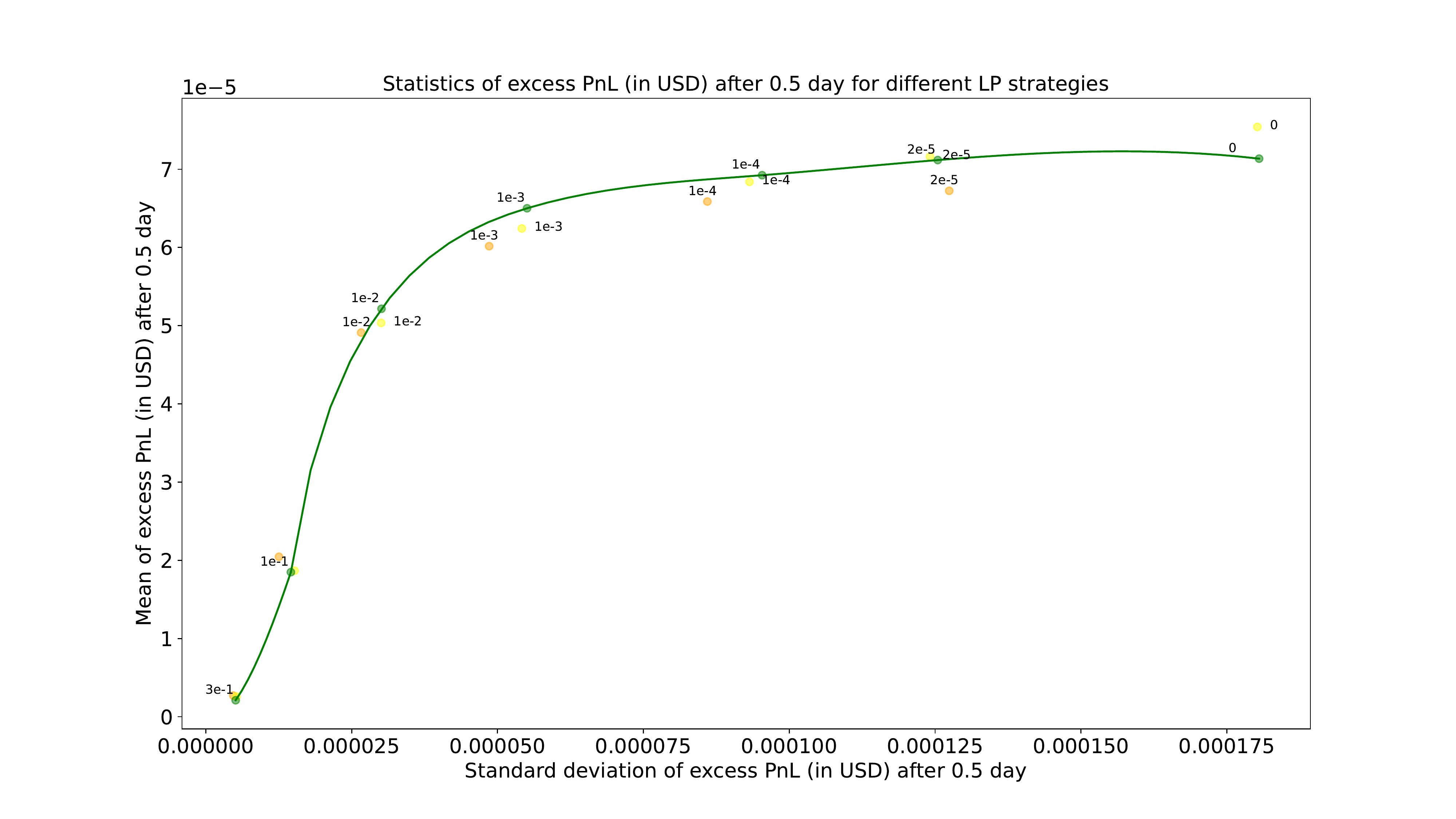}\\
\caption{Performance of the previous optimal strategy in terms of mean / standard deviation of excess PnL. In green: efficient frontier, obtained by playing the optimal strategy for different levels of risk aversion with complete information. In yellow: performance of the same optimal strategy for different levels of risk aversion with a lagged oracle. In orange: performance of the same optimal strategy for different levels of risk aversion with a lagged oracle and with arbitrage flow. The number next to the point corresponds to the value of $\gamma$.}
\label{pic_3}
\end{figure}

\begin{figure}[h!]
\centering
\includegraphics[width=0.95\textwidth]{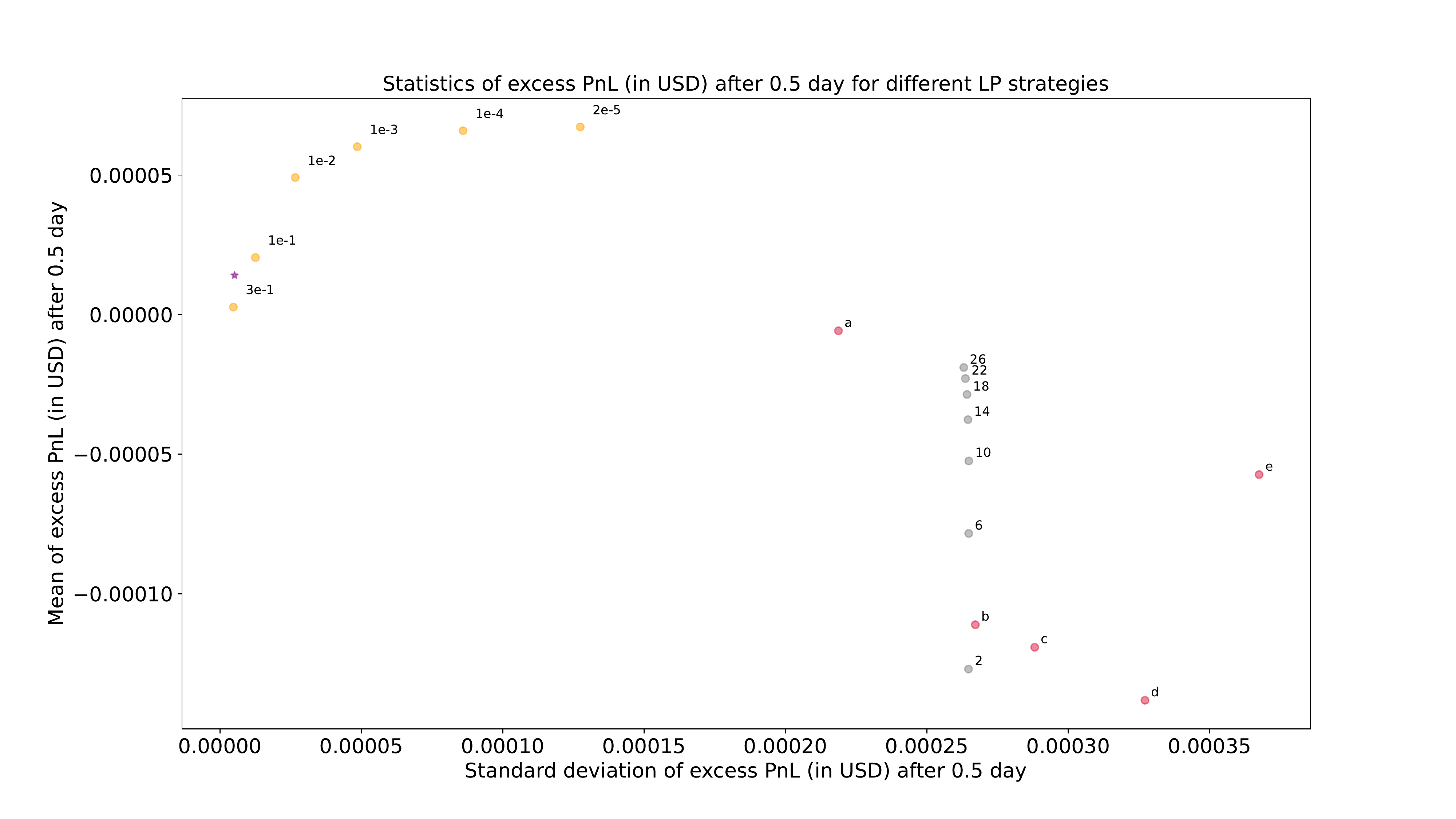}\\
\caption{Performance of strategies in terms of mean / standard deviation of excess PnL with a lagged oracle and with arbitrage flow. In grey: CPMM with fees. In pink: CFMM without market exchange rate oracle as described in \cite{curve_2019, curve_2021} for different sets of realistic parameters. In purple ($\star$): AMM with market exchange rate oracle as described in \cite{bouba_2021}. In orange: performance of the efficient strategies for different levels of risk aversion.}
\label{pic_4}
\end{figure}

Partial information regarding the market exchange rate can sometimes result in arbitrage opportunities for LTs. These arbitrage opportunities are already taken into account in the demand curves modeled by the intensity functions, though not in a systematic way. In other words, if a price appears to be very good for LTs, the probability that a transaction occurs is very high in our model. In practice however, there exists a category of agents who systematically exploit arbitrage opportunities: they trade with the AMM until arbitrage opportunities disappear.\footnote{Of course, these trades come with a cost for arbitrageurs, who still have to pay the fees associated with a swap on an other (centralized) exchange. In our case, we assume a proportional cost of $7.5 \text{ bps}$. Moreover, note that in our simulations, arbitrageurs trade a size that is optimal for them -- and thus does not necessarily correspond to the size of the other trades.} The above analysis regarding market exchange rate oracles can then only be complete if we take into account these arbitrageurs. In practice, arbitrageurs can represent a significant volume as soon as the market exchange rate moves outside of the spread offered by the AMM, and we compare in Figure \ref{pic_3} the performance of the efficient strategies in the presence of a 10-second delayed oracle and with arbitrageurs. We also provide in Figure \ref{pic_4} a complete risk / return analysis of the different strategies studied in this paper, in the presence of arbitrageurs and incomplete information (for those relying on oracles). \\

What we observe in Figure \ref{pic_3} is that the presence of arbitrageurs who systematically and continuously keep the offered exchange rate in a narrow range around the market exchange rate tends to increase the expected excess PnL and reduce risk. This may sound counter-intuitive, but it comes from the fact that those arbitrageurs actually ‘‘protect’’ the AMM against trades with traditional LTs at prices even further away from the market exchange rate (this could be related to the findings of \cite{milionis2023automated2}). Of course, the efficient frontier documented in Figure \ref{pic_3} does not take into account this additional flow. Building a model that would internalize both delayed oracles and (systematic) arbitrageurs and allow to derive optimal strategies in this context remains an open problem. \\

Figure \ref{pic_4} confirms that the use of a market exchange rate oracle, even delayed, really makes a difference in terms of risk / return profile. Nevertheless, it is important to note that introducing an oracle creates a fundamentally different protocol which relies on external data. One of the issues with oracles (as noted in \cite{tjiam2021oracle} and then in \cite{mackinga2022twap}), is that they could be manipulated. Using an oracle in an AMM protocol should therefore be performed carefully in order for LPs to really achieve the promised improved risk-adjusted performance (for a detailed explanation of the next generation of decentralized oracles see \cite{breidenbach2021chainlink}).\\

\section{Conclusion}
\label{concl}

In this paper, we provide an analysis of AMMs in a mean / standard deviation framework inspired by both Markowitz' modern portfolio theory and the recent literature on optimal market making. We show that traditional CFMMs (including CPMMs) with different levels of transaction fees perform poorly relative to the theoretical efficient frontier and very often exhibit negative excess PnL. We also show that allowing an AMM to get information about the current market exchange rate (through an oracle) can significantly improve performance. Such an oracle-based AMM would quote a bid / ask spread around the current market exchange rate based on its reserves. This design significantly reduces the volatility of the excess PnL while delivering a positive excess PnL on average. Our results are robust to the presence of a lagged oracle and to the introduction of adverse selection by LTs and arbitrageurs. Nevertheless, while introducing an oracle in the AMM design can significantly improve the risk-adjusted performance of LPs, it comes at the cost that the oracle itself should be carefully designed to avoid introducing additional attack~vectors.

\section*{Statement and acknowledgment}

The research carried out for this paper benefited from the support of the Research Program ‘‘Decentralized Finance and Automated Market Makers’’, a Research Program under the aegis of Institut Europlace de Finance, in partnership with Apesu / Swaap Labs.\\

The content of this article has been presented at several conferences and seminars: BlockSem seminar (Paris, France), FFEA 2nd Spring Workshop on FinTech (Ghent, Belgium), The 3rd workshop on Decentralized Finance (Bol, Croatia), DeFOx seminar (Oxford, UK), Euro Working Group on Commodities and Financial Modelling meeting (Rome, Italy), SIAM Conference on Financial Mathematics and Engineering (Philadelphia, USA), AMaMeF conference (Bielefeld, Germany), Florence-Paris Workshop on Mathematical Finance (Florence, Italy), Frontiers in Quantitative Finance Seminar (London, UK), Séminaire Bachelier (Paris, France), 16th Financial Risks International Forum (Paris, France), London-Paris Bachelier Workshop on Mathematical Finance (London, UK), Blockchain@X-OMI Workshop on Blockchain and Decentralized Finance (Palaiseau, France), Apéro DeFi (Paris, France). The discussions that took place during these events have substantially contributed to improving the presentation of our results.\\

We extend our sincere gratitude to the two anonymous referees for their insights and constructive feedback, which greatly enhanced the quality of our article.\\

\nocite{*}
\bibliographystyle{plain}

\end{document}